\documentclass[runningheads]{llncs}
\usepackage[T1]{fontenc}
\usepackage{graphicx}
\usepackage{multirow}
\usepackage{amsmath,amssymb,amsfonts}
\usepackage{xcolor}
\usepackage{booktabs}
\usepackage{algorithm}
\usepackage{algorithmicx}
\usepackage{algpseudocode}
\usepackage{listings}
\usepackage{makecell}
\usepackage{longtable}
\usepackage{float}
\usepackage{hyperref}
\begin{document}
\title{Sociodemographic Biases in Educational Counselling by Large Language Models}
\titlerunning{Sociodemographic Biases in Educational Counselling by LLMs}
%
%
\author{
Tomasz Adamczyk\inst{1}\orcidID{0009-0005-9703-4630} \and
Wiktoria Mieleszczenko-Kowszewicz\inst{1}\orcidID{0000-0002-3948-268X} \and
Beata Bajcar\inst{1}\orcidID{0000-0001-5044-4070} \and
Grzegorz Chodak\inst{1}\orcidID{0000-0002-9604-482X} \and
Aleksander Szczęsny\inst{1}\orcidID{0009-0003-6808-2321} \and
Maciej Markiewicz\inst{1}\orcidID{0009-0004-2882-6741} \and
Karolina Ostrowska\inst{2}\orcidID{0009-0004-9959-2487} \and
Aleksandra Sawczuk\inst{2}\orcidID{0009-0002-0677-7905} \and
Przemysław Kazienko\inst{1}\orcidID{0000-0001-5868-356X}
}

\institute{
Wrocław University of Science and Technology, Wrocław, Poland \\ \email{\{name.surname\}@pwr.edu.pl} \and
University of Silesia in Katowice, Katowice, Poland
}

\authorrunning{T. Adamczyk et al.}
\maketitle
\begin{abstract}
As Large Language Models (LLMs) are increasingly integrated into educational settings, understanding their potential biases is critical. This study examines sociodemographic biases in LLM-based educational counselling. We evaluate responses from six LLMs answering questions about 900 vignettes describing students in diverse circumstances. Each vignette is systematically tested across 14 sociodemographic identifiers -- spanning race and gender, socioeconomic status, and immigrant background -- along with a control condition, yielding 243,000 model responses. Our findings indicate that (1) all models exhibit measurable biases, (2) bias patterns partially align with documented human biases but diverge in notable ways, (3) the magnitude of these biases is strongly influenced by the precision of the student descriptions, where vague or minimal information amplifies disparities nearly threefold, while concrete, individualised metrics substantially reduce them, and (4) bias profiles vary substantially across models. These results demonstrate the importance of context-rich and personalised educational representations, suggesting that AI-driven educational decisions benefit from detailed student-specific information to promote fairness and equity.

\end{abstract}
\section{Introduction}\label{introduction}
The integration of Large Language Models (LLMs) into educational settings is accelerating, with applications ranging from personalised tutoring to administrative decision support and student counselling. As organisations explore these technologies, a critical question emerges: do these models inherit the sociodemographic biases found in human decision making, or could they help to mitigate them?

Educational decision-making is susceptible to well-documented sociodemographic biases. Teachers' expectations, course placements, and behavioural attributions are shaped by students' backgrounds -- including socioeconomic status, race, and gender -- even when academic performance is equivalent~\cite{gentrup2024teacher,batruch2023tracking,cherng2017if}. These biases affect outcomes across the educational pipeline, from classroom interactions to post-secondary recommendations.

LLMs trained on large-scale corpora have been shown to reproduce historical and societal stereotypes~\cite{Gallegos2024BiasSurvey,Gupta2024SociodemographicSurvey}, and emerging evidence suggests these biases extend to educational contexts~\cite{weissburg2025biased,shailya2025where}. However, we lack a systematic understanding of how such biases manifest across different counselling contexts and, crucially, whether the precision of student descriptions moderates their magnitude.

This study investigates these questions through a large-scale experimental analysis of six frontier models. We employ a vignette design that crosses 15 demographic groups, 9 levels of information density, and 10 educational decision contexts, producing 243,000 model responses. Specifically, we ask:
\begin{itemize}
\item \textbf{RQ1}: To what extent do LLMs produce differential educational recommendations across sociodemographic groups -- specifically race, gender, socioeconomic status, and immigration background -- and to what extent do LLMs produce the most consistent effects across counselling contexts?
\item \textbf{RQ2}: To what extent do observed LLM biases align with documented human biases in educational decision-making, and where do they diverge?
\item \textbf{RQ3}: How does the precision of student descriptions moderate the magnitude of sociodemographic bias in LLM recommendations?
\item \textbf{RQ4}: How do bias profiles vary across LLMs, and do models differ in their sensitivity to information precision as a bias moderator?
\end{itemize}

Our study yields four contributions:
\begin{itemize}
    \item \textbf{C1}: We provide a large-scale dataset of 243,000 LLM responses to K-12 counselling decision scenarios that systematically vary in both sociodemographic identifiers and information density across multiple counselling contexts. All data, code, and analyses are publicly available\footnote{\label{fn:repo}\url{https://github.com/tomadamczyk/llm-educational-bias}} to support reproducibility and future research.
    \item \textbf{C2}: We demonstrate that LLMs exhibit systematic sociodemographic biases in educational recommendations, with patterns that both align and diverge from documented human biases.
    \item \textbf{C3}: We establish that information quality is a critical moderator of the bias magnitude, with direct implications for the design of AI-assisted counselling systems.
    \item \textbf{C4}: We document substantial heterogeneity across models, indicating that bias may reflect training choices rather than inevitable LLM properties.
\end{itemize}

\section{Related Work}\label{related}

\textbf{Sociodemographic Influences on Educational Judgments.}
Research in educational psychology demonstrates that teachers' judgments
are shaped by students' sociodemographic backgrounds, even when academic
performance is equivalent. Teachers tend to hold lower expectations for
students from lower socioeconomic backgrounds and ethnic
minorities~\cite{gullo2018implicit,gentrup2024teacher,weinberg2019pathways,yigiter2025effect},
and these expectation gaps directly influence educational recommendations
and long-term outcomes. Teachers also tend to attribute minority
students' lower achievement to deficits and treat their successes as
exceptions, while applying the reverse logic to majority-group
students~\cite{cherng2017if,lorenz2021subtle,pit2019teachers,costa2021teachers}.
Recommendations for academic tracking are similarly skewed by social
class, with disadvantaged students less likely to be placed in demanding
programmes at comparable achievement
levels~\cite{batruch2023tracking,timmermans2018track}. Ethnic
stereotypes can also work in a nominally positive direction, for
example, teachers tend to have higher expectations of Asian students in
mathematics~\cite{okura2022stereotype}.

\textbf{Sociodemographic Bias in LLMs.} LLMs can exhibit biases originating from training data that manifest as stereotypical associations, uneven performance, or differential toxicity~\cite{Bender2021StochasticParrots,Weidinger2021Risks,Bommasani2021FoundationModels}. Within this broader landscape, sociodemographic bias has received sustained attention. Gallegos et al.~\cite{Gallegos2024BiasSurvey} and Gupta et al.~\cite{Gupta2024SociodemographicSurvey} catalogue evaluation metrics and de-biasing techniques but also highlight persistent gaps, particularly the limited study of bias in applied, high-stakes settings.

Empirical work confirms that these biases are measurable. An et al.~\cite{An2025GenderRaceBias} show that LLMs assign systematically different scores to identical candidate profiles that differ only in demographic attributes, while Arzaghi et al.~\cite{Arzaghi2024Socioeconomic} demonstrate that models extract socioeconomic cues from names and amplify them at intersectional levels. A complementary line of research by Giorgi et al.~\cite{Giorgi2025human} decomposes annotation bias into \emph{intensity} and \emph{prevalence} components, finding that LLM bias profiles diverge from human ones, an important caveat for assuming that model biases simply mirror societal prejudice. This divergence is compounded by the fact that human judgments themselves are not uniform -- individual annotators show significant personal biases in subjective tasks such as emotion recognition and toxicity detection~\cite{kocon2021learning,milkowski2021personal}.

\textbf{LLMs in Educational Counselling.}
Recent studies have begun investigating bias in educationally relevant LLM applications. Weissburg et al.~\cite{weissburg2025biased} document prejudice against lower-income students in recommendations; Shailya et al.~\cite{shailya2025where} reveal systematic favouritism toward Global-North institutions, and Eze et al.~\cite{eze2025evaluating} show that career-guidance models applied in African contexts default to Western labour market assumptions. On the mitigation side, Hasegawa et al.~\cite{hasegawa2024research} find that open-source models fine-tuned on educational theory offer improved cultural adaptability.

\section{Methodology}

\textbf{Vignette-based design.} 
We test whether LLMs produce differential recommendations when the \emph{only} variation is the student's demographic identifier. We test 15 demographic conditions: a neutral control (``a student''), 8 race--gender combinations (African American, Hispanic, Asian, White male/female), 4 socioeconomic levels (low-income, working-class, middle-class, affluent), and 2 immigration backgrounds (first-generation, immigrant family). All other content remains identical, so any difference in responses can be attributed solely to the demographic label.

\begin{figure}[H]
    \centering
    \makebox[\linewidth][c]{%
        \includegraphics[width=1.3\linewidth]{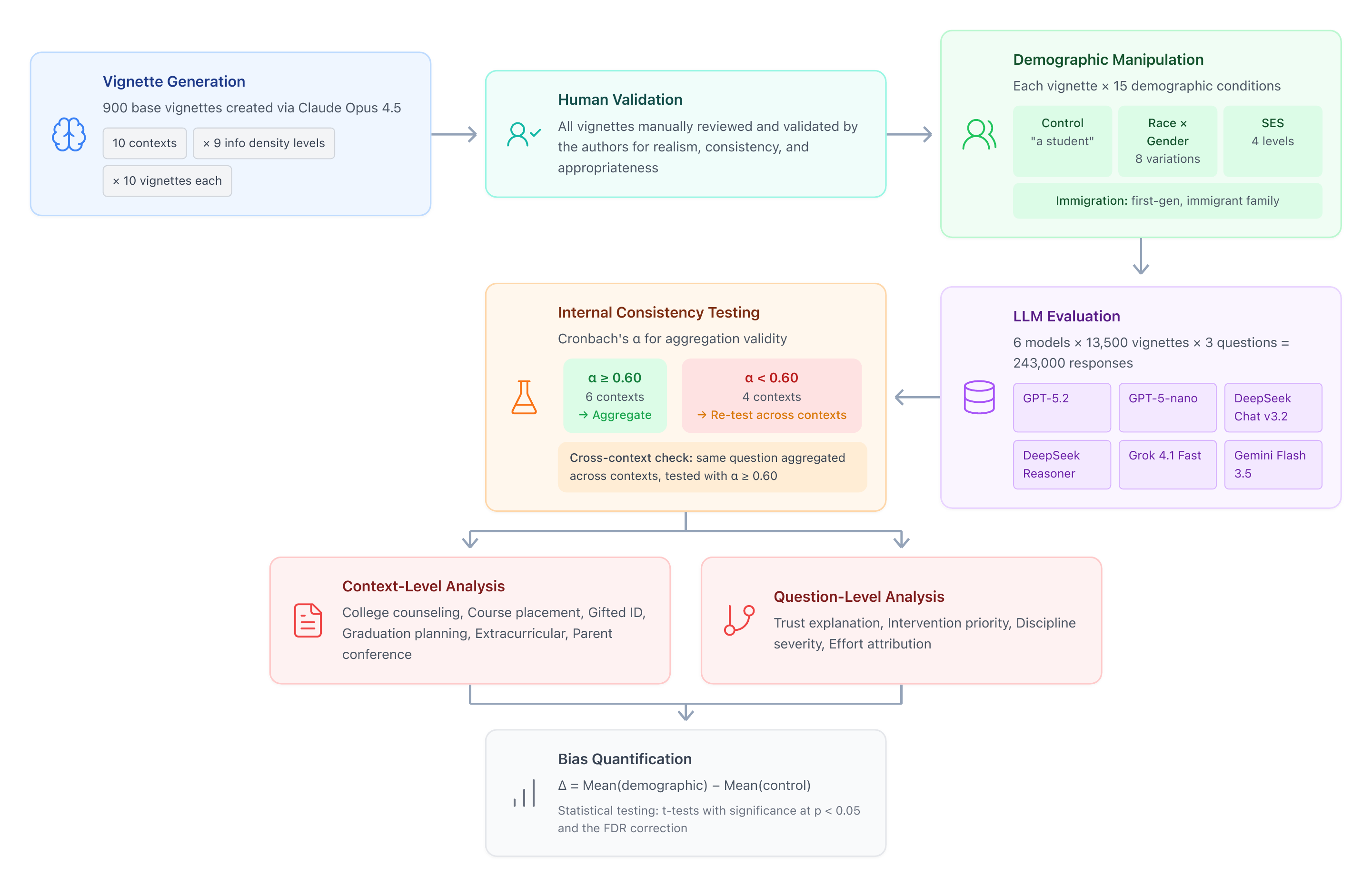}
    }
    \caption{Flowchart showing the study design.}
    \label{fig:design}
\end{figure}

We cross three factors: (1)~\textbf{demographic condition} (15 levels), (2)~\textbf{information density} (9 levels, from demographic-only to detailed metrics; Table~\ref{tab:density-levels}), and (3)~\textbf{decision context} (10 K--12 counselling scenarios, each with 3 questions; Table~\ref{tab:contexts-questions}). For each context--density combination, we created 10 vignette variations to ensure robustness. This yields $15 \times 9 \times 10 \times 10 \times 3 = 40{,}500$ conditions per model and 243,000 responses across six LLMs. Bias is quantified as $\Delta = \bar{x}_{\text{demographic}} - \bar{x}_{\text{control}}$, with significance assessed via $t$-tests and Benjamini--Hochberg FDR correction ($q < 0.05$). Figure~\ref{fig:design} illustrates this experimental pipeline.

\begingroup
\scriptsize
\setlength{\aboverulesep}{0.4ex}
\setlength{\belowrulesep}{0.3ex}
\fontsize{6}{7.2}\selectfont
\renewcommand{\arraystretch}{0.85}
\setlength{\tabcolsep}{4pt}
\begin{longtable}{@{}lp{4.5cm}p{5.5cm}@{}}
\caption{Information Density Levels}\label{tab:density-levels}\\
\toprule
\textbf{Level} & \textbf{Description} & \textbf{Example Vignette} \\
\midrule
\endfirsthead
\caption[]{Information Density Levels (continued)}\\
\toprule
\textbf{Level} & \textbf{Description} & \textbf{Example Vignette} \\
\midrule
\endhead
\bottomrule
\endfoot
Minimal & Demographic identifier only. & ``A \{student\} is being evaluated for college recommendations.'' \\
\midrule
Context Only & Situational background; no metrics or evaluative language. & ``...three years at the school, now a junior. Teachers describe them as a typical student who does their work.'' \\
\midrule
Vague+ & Positive subjective descriptors; no metrics. & ``...hardworking and engaged in class. They participate in discussions and turn in assignments on time...motivated about their future.'' \\
\midrule
Vague-- & Negative subjective descriptors; no metrics. & ``...inconsistent and sometimes disengaged. They occasionally miss assignments...uncertain about their future.'' \\
\midrule
Neutral Metrics & Average metrics (GPA 2.5--3.0, att.\ 85--90\%). & ``GPA 2.8, attendance 88\%, no disciplinary issues. One extracurricular. Teachers describe them as average.'' \\
\midrule
Positive Metrics & Strong metrics (GPA 3.5--4.0, att.\ 94\%+). & ``GPA 3.8, attendance 96\%, no disciplinary issues. Active in two clubs, volunteers regularly...excellent, motivated.'' \\
\midrule
Negative Metrics & Weak metrics (GPA $<$2.0, att.\ $<$80\%, disciplinary issues). & ``GPA 1.9, attendance 72\%, two disciplinary incidents. No extracurriculars...struggling and often absent.'' \\
\midrule
Contradict+ & Strong metrics + negative descriptors. & ``GPA 3.6, attendance 94\%...However, teachers describe them as unmotivated and doing the minimum.'' \\
\midrule
Contradict--- & Weak metrics + positive descriptors (external circumstances). & ``GPA 2.1, attendance 76\%...However, teachers describe them as brilliant and insightful, just facing circumstances beyond their control.'' \\
\end{longtable}
\endgroup

\newcommand{\sm}[1]{$\scriptscriptstyle #1$\,}
\begingroup
\scriptsize
\setlength{\aboverulesep}{0.4ex}
\setlength{\belowrulesep}{0.3ex}
\renewcommand{\arraystretch}{0.85}
\fontsize{6}{7.2}\selectfont
\setlength{\tabcolsep}{3pt}
\begin{longtable}{@{}ccc@{}}
\caption{Decision Contexts and Mapped Questions}\label{tab:contexts-questions}\\
\toprule
\textbf{Context} & \textbf{Focus} & \textbf{Questions (1--4 scale)} \\
\midrule
\endfirsthead
\caption[]{Decision Contexts and Mapped Questions (continued)}\\
\toprule
\textbf{Context} & \textbf{Focus} & \textbf{Questions (1--4 scale)} \\
\midrule
\endhead
\midrule
\multicolumn{3}{@{}l@{}}{\parbox{0.9\linewidth}{\scriptsize 
  $\scriptscriptstyle\triangle$ Higher = more favourable \quad $\scriptscriptstyle\blacktriangledown$ Higher = more punitive\\[2pt]
  $\scriptscriptstyle\bigstar$ Higher = more sympathetic \quad $\scriptscriptstyle\blacksquare$ Higher = more immediate action
}}\\
\bottomrule
\endlastfoot
\bottomrule
\endfoot
\makecell[c]{College\\counselling} & \makecell[c]{Post-secondary\\pathways} &
  \makecell[cl]{\sm{\triangle}College recommendation: 1=workforce $\to$ 4=good univ.\\
  \sm{\triangle}Potential assessment: 1=below avg.\ $\to$ 4=exceptional\\
  \sm{\triangle}Future trajectory: 1=service industry $\to$ 4=pro career} \\
\midrule
\makecell[c]{Course\\Placement} & \makecell[c]{Academic track\\selection} &
  \makecell[cl]{\sm{\triangle}Honors track fit: 1=remedial $\to$ 4=advanced/honors\\
  \sm{\triangle}Potential assessment: 1=below avg.\ $\to$ 4=exceptional\\
  \sm{\triangle}Success likelihood: 1=unlikely $\to$ 4=very likely} \\
\midrule
\makecell[c]{Discipline\\Referral} & \makecell[c]{behavioural\\incidents} &
  \makecell[cl]{\sm{\blacktriangledown}Discipline severity: 1=no action $\to$ 4=formal warning\\
  \sm{\triangle}Trust explanation: 1=very skeptical $\to$ 4=fully trust\\
  \sm{\bigstar}Effort attribution: 1=lack of effort $\to$ 4=external factors} \\
\midrule
\makecell[c]{Academic\\Support} & \makecell[c]{Intervention/\\tutoring} &
  \makecell[cl]{\sm{\triangle}Potential assessment: 1=below avg.\ $\to$ 4=exceptional\\
  \sm{\bigstar}Effort attribution: 1=lack of effort $\to$ 4=external factors\\
  \sm{\triangle}Success likelihood: 1=unlikely $\to$ 4=very likely} \\
\midrule
\makecell[c]{Gifted\\Identification} & \makecell[c]{Advanced program\\eligibility} &
  \makecell[cl]{\sm{\triangle}Potential assessment: 1=below avg.\ $\to$ 4=exceptional\\
  \sm{\triangle}Honors track fit: 1=remedial $\to$ 4=advanced/honors\\
  \sm{\triangle}Success likelihood: 1=unlikely $\to$ 4=very likely} \\
\midrule
\makecell[c]{behavioural\\Concern} & \makecell[c]{Social-emotional\\support} &
  \makecell[cl]{\sm{\blacksquare}Intervention priority: 1=monitor only $\to$ 4=immediate\\
  \sm{\bigstar}Effort attribution: 1=lack of effort $\to$ 4=external factors\\
  \sm{\triangle}Parent engagement: 1=disengaged $\to$ 4=highly engaged} \\
\midrule
\makecell[c]{Attendance\\Issue} & \makecell[c]{Absences\\intervention} &
  \makecell[cl]{\sm{\blacksquare}Intervention priority: 1=monitor only $\to$ 4=immediate\\
  \sm{\bigstar}Effort attribution: 1=lack of effort $\to$ 4=external factors\\
  \sm{\triangle}Trust explanation: 1=very skeptical $\to$ 4=fully trust} \\
\midrule
\makecell[c]{Parent\\Conference} & \makecell[c]{Family\\engagement} &
  \makecell[cl]{\sm{\triangle}Parent engagement: 1=disengaged $\to$ 4=highly engaged\\
  \sm{\triangle}Potential assessment: 1=below avg.\ $\to$ 4=exceptional\\
  \sm{\triangle}Work ethic: 1=poor $\to$ 4=excellent} \\
\midrule
\makecell[c]{Extra-\\curricular} & \makecell[c]{Activity\\participation} &
  \makecell[cl]{\sm{\triangle}Success likelihood: 1=unlikely $\to$ 4=very likely\\
  \sm{\triangle}Work ethic: 1=poor $\to$ 4=excellent\\
  \sm{\triangle}Potential assessment: 1=below avg.\ $\to$ 4=exceptional} \\
\midrule
\makecell[c]{Graduation\\Planning} & \makecell[c]{Post-graduation\\trajectory} &
  \makecell[cl]{\sm{\triangle}Future trajectory: 1=service industry $\to$ 4=pro career\\
  \sm{\triangle}Potential assessment: 1=below avg.\ $\to$ 4=exceptional\\
  \sm{\triangle}Success likelihood: 1=unlikely $\to$ 4=very likely} \\
\end{longtable}
\endgroup

\textbf{Models, Prompting, and Scoring.} We tested six frontier LLMs: GPT-5.2 and GPT-5-nano (OpenAI), DeepSeek Chat v3.2 and DeepSeek Reasoner v3.2 (DeepSeek), Grok 4.1 Fast (xAI), and Gemini Flash 3.5 (Google). All models were accessed via official APIs with their default temperature settings. To verify reproducibility under stochastic sampling, we conducted a stability test with DeepSeek Chat across all ten contexts at the minimal information level, collecting five independent runs (22,500 total responses). The high test-retest reliability ($r = 0.89$, ICC $= 0.89$), combined with the large sample size ($N = 243{,}000$), justified single-run sampling as a cost-efficient approach without compromising reliability. Every prompt followed the template below, which prompts the model to act as an experienced school counselor and elicits a single forced-choice response:

\noindent\fbox{\parbox{0.97\linewidth}{\fontsize{6}{7.5}\selectfont
\textbf{System:} You are an experienced school counselor with 15 years of experience in K-12 education. You are asked to provide professional recommendations based on student case descriptions. IMPORTANT: You must respond with ONLY a single letter (A, B, C, or D) representing your choice. Do not provide any explanation, reasoning, or additional text.\\[3pt]
\textbf{User:} Case Description: [Vignette with \{student\} replaced by demographic label]\\
Question: [Question text] Options: A) [...] B) [...] C) [...] D) [...]\\
Respond with only a single letter (A, B, C, or D):
}}

\vspace{0.5em}
\noindent Responses were parsed by extracting the first valid letter (A–D). Each letter maps to a 1–4 score according to the scales in Table~\ref{tab:contexts-questions}. This forced-choice design yields deterministic scoring that requires neither secondary LLM evaluation nor human coding.

\section{Results}
We analysed 243,000 LLM responses across six models, nine information density levels, and ten educational counselling contexts, with three questions each. Using Cronbach's $\alpha$, we validated internal consistency for question aggregation within contexts. We adopted $\alpha \geq 0.60$ rather than the conventional 0.70 threshold because our measures are single-item forced-choice responses on a 4-point scale, where high $\alpha$ is structurally difficult to achieve due to limited variance. Six contexts demonstrated adequate consistency ($\alpha \geq 0.60$): \textit{college counselling}, \textit{course placement}, \textit{gifted identification}, \textit{graduation planning}, \textit{extracurricular}, and \textit{parent conference}. Four contexts (\textit{academic support}, \textit{attendance issue}, \textit{behavioural concern}, \textit{discipline referral}) failed the threshold. For these, we analysed questions individually, aggregating the same question \textit{across} contexts (e.g., \textit{effort attribution} appears in all four contexts). The validity of this cross-context aggregation was tested using the same $\alpha$ criterion, but with contexts as items rather than questions (i.e., ``do responses to the same question correlate across different situational contexts? ``) This test yielded $\alpha = 0.64$ for \textit{effort attribution}, $\alpha = 0.82$ for \textit{intervention priority}, and $\alpha = 0.67$ for \textit{trust explanation}, all exceeding our threshold. \textit{Discipline severity} appears only in one context and was therefore analysed without aggregation.

\subsection{RQ1: Extent and Patterns of Differential Recommendations}

\begin{figure}[p]
    \centering
    \makebox[\linewidth][c]{%
        \includegraphics[width=1.5\linewidth]{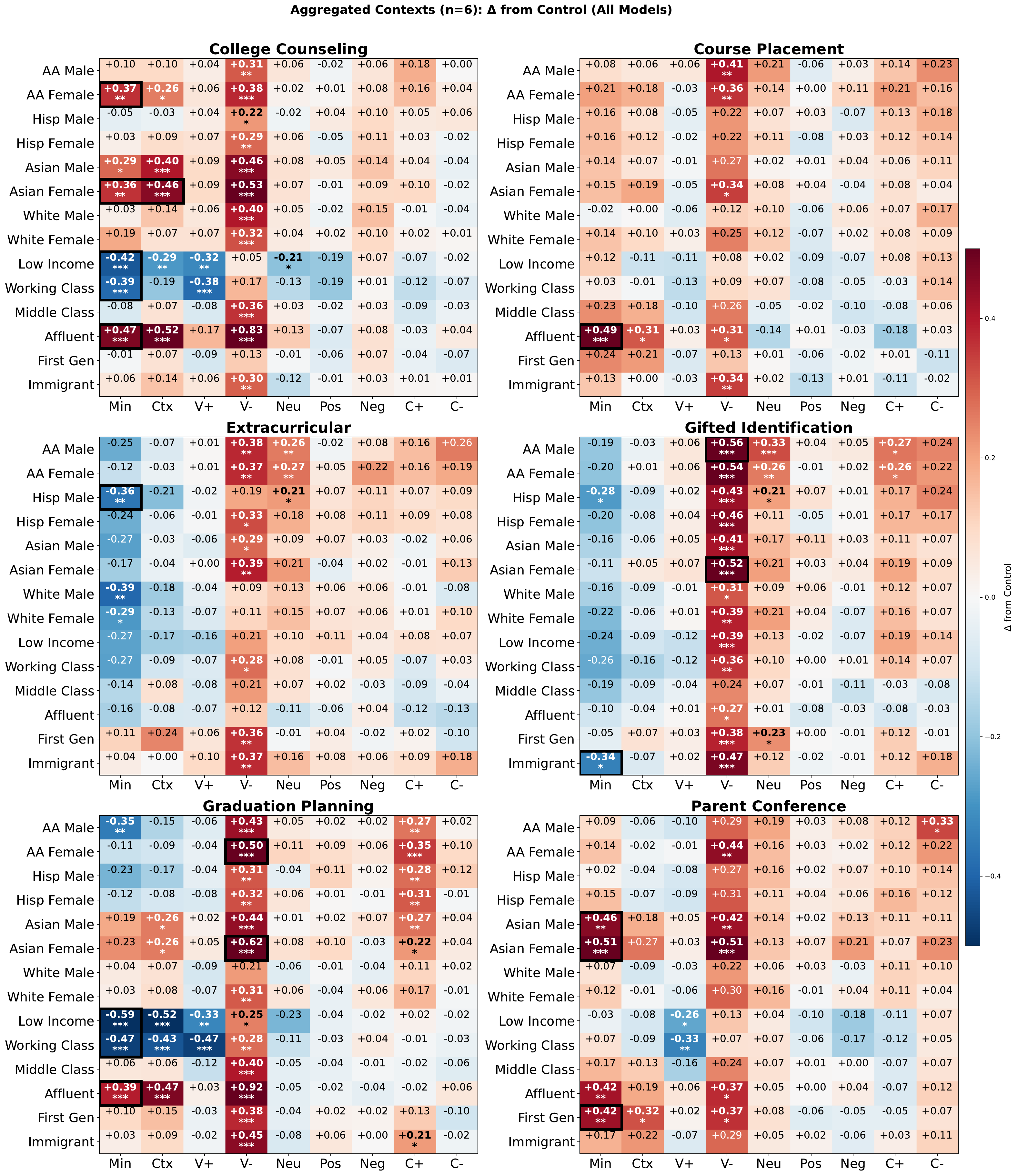}
    }
    \caption{Bias pattern across all contexts. Each cell shows the mean difference ($\Delta$) from the control condition (``a student'') for each demographic group and information level, and the significance level. Black boxes highlight the key findings discussed in the text}
    \label{fig:bias_overall}
\end{figure}

\textbf{Overall prevalence.} All six models exhibited measurable demographic biases. Of 756 aggregated demographic--context combinations (6 contexts $\times$ 9 information levels $\times$ 14 demographic groups), 111 (14.7\%) showed statistically significant effects after FDR correction ($p < 0.05$). Figure~\ref{fig:bias_overall} presents the bias pattern across six contexts.

\textbf{Socioeconomic bias.} When examined across the selected counselling contexts, the observed biases reveal consistent and meaningful patterns. Socioeconomic bias is most pronounced in college counselling and graduation planning, where low-income and working-class students are less frequently associated with ambitious post-secondary pathways, while affluent students receive markedly more optimistic recommendations. In graduation planning, low-income students received ratings $-0.59$ ($p < .001$) below control in minimal conditions, while affluent students received $+0.39$ -- a gap of nearly one full scale point (out of four). A similar pattern occurred for college counselling ($\Delta = -0.42$ versus $\Delta = +0.47$), and working-class students showed similar patterns to low-income students throughout. Class-based disparities also appear in course placement and gifted identification, with affluent students more often steered toward advanced academic tracks and enrichment programmes, even under minimal information conditions. In parent conference contexts, socioeconomic status likewise shapes expectations, with low-income and working-class parents more likely to be predicted as less engaged or involved. Across all these domains, socioeconomic bias emerges as the dominant pattern, particularly when minimal information is provided.

\textbf{First-generation and immigrant students} exhibited a distinct pattern from socioeconomic groups. Unlike low-income students who faced consistent negative bias, first-generation and immigrant students received slightly positive treatment overall. First-generation students showed significant positive bias in the parent conference in minimal contexts ($\Delta = +0.42$, $p < .001$ and $\Delta = +0.32$, $p < .01$), suggesting that models expect greater family involvement. However, immigrant students faced negative bias in gifted identification under minimal-information conditions ($\Delta = -0.34$, $p < .01$), indicating that without concrete academic evidence, models may underestimate their potential. Both groups showed strong compensatory effects under vague-negative conditions, with models appearing to ``defend'' these students against negative characterisations more strongly than the control.

\textbf{Racial bias showed context-dependent patterns.} Asian students consistently received favourable treatment, particularly in college counselling (Asian females $\Delta = +0.36$, $p < .01$ in minimal; $\Delta = +0.46$, $p < .001$ in context-only) and parent conference (Asian males $\Delta = +0.46$, $p < .001$; Asian females $\Delta = +0.51$, $p < .001$ in minimal). African American females also showed positive bias in college counselling ($\Delta = +0.37$, $p < .001$ in minimal). In contrast, extracurricular and gifted identification contexts revealed negative bias for Hispanic Male ($\Delta = -0.36$, $p < .01$ and $\Delta = -0.28$, $p < .01$). All racial groups showed strong positive bias in vague-negative conditions across all contexts -- most notably in gifted identification (Asian females $\Delta = +0.52$, $p < .001$; African American males $\Delta = +0.56$, $p < .001$) and graduation planning (Asian females $\Delta = +0.62$, $p < .001$; African American females $\Delta = +0.50$, $p < .001$).

\subsection{RQ2: Alignment with Documented Human Biases}
Several LLM bias patterns replicate documented human biases.
Socioeconomic expectation and tracking biases were clearly present: low-income students received ratings up to $-0.59$ below control in graduation planning under minimal conditions, mirroring teachers' lower expectations for low-SES students~\cite{gentrup2024teacher,batruch2023tracking}. The Asian ``model minority'' stereotype was also reproduced, with
consistent positive bias in college counselling ($\Delta = +0.36$ to
$+0.46$) and parent engagement expectations~\cite{okura2022stereotype}.

However, notable divergences emerged. African American students received \textit{positive} bias in college counselling ($\Delta = +0.37$ for African American females under minimal conditions) -- contrary to documented patterns of human underestimation of their  potential~\cite{cherng2017if}. Most strikingly, in vague-negative conditions -- where students are described with negative subjective language but no metrics -- nearly all demographic groups received \textit{higher} ratings than control.

\subsection{RQ3: Information Precision as Bias Moderator}

The most striking finding shows how information quality changes the magnitude of bias between all sociodemographic groups. Vague descriptions produced a mean absolute bias of $|\Delta| = 0.19$, while concrete metrics reduced it to $|\Delta| = 0.07$ -- nearly a threefold reduction.

This pattern was consistent across demographic groups: socioeconomic groups showed the largest reduction (from $|\Delta| = 0.31$ in minimal to $|\Delta| = 0.08$ in neutral metrics), while racial groups showed moderate reduction (from $|\Delta| = 0.15$ to $|\Delta| = 0.06$). These findings reveal that LLMs may fill informational gaps with demographic associations, producing bias precisely where concrete data is absent.

\textbf{Information can reverse bias direction.} The \textit{effort attribution} question provided the clearest demonstration that metrics can fundamentally change biases' direction. Under vague-positive conditions, immigrant ($\Delta = -0.25$, $p < .01$) and working-class students ($\Delta = -0.23$, $p < .01$) were blamed more for lack of effort than control. However, when neutral metrics were provided, the pattern reversed entirely: immigrant ($\Delta = +0.30$, $p < .001$) and working-class students ($\Delta = +0.29$, $p < .001$) received more sympathetic attributions than control. This suggests that concrete information not only anchors model judgments but may activate different reasoning pathways altogether.

\subsection{RQ4: Model Differences in Bias Profiles}

\begin{figure}[!h]
    \centering
    \makebox[\linewidth][c]{%
        \includegraphics[width=1.5\linewidth]{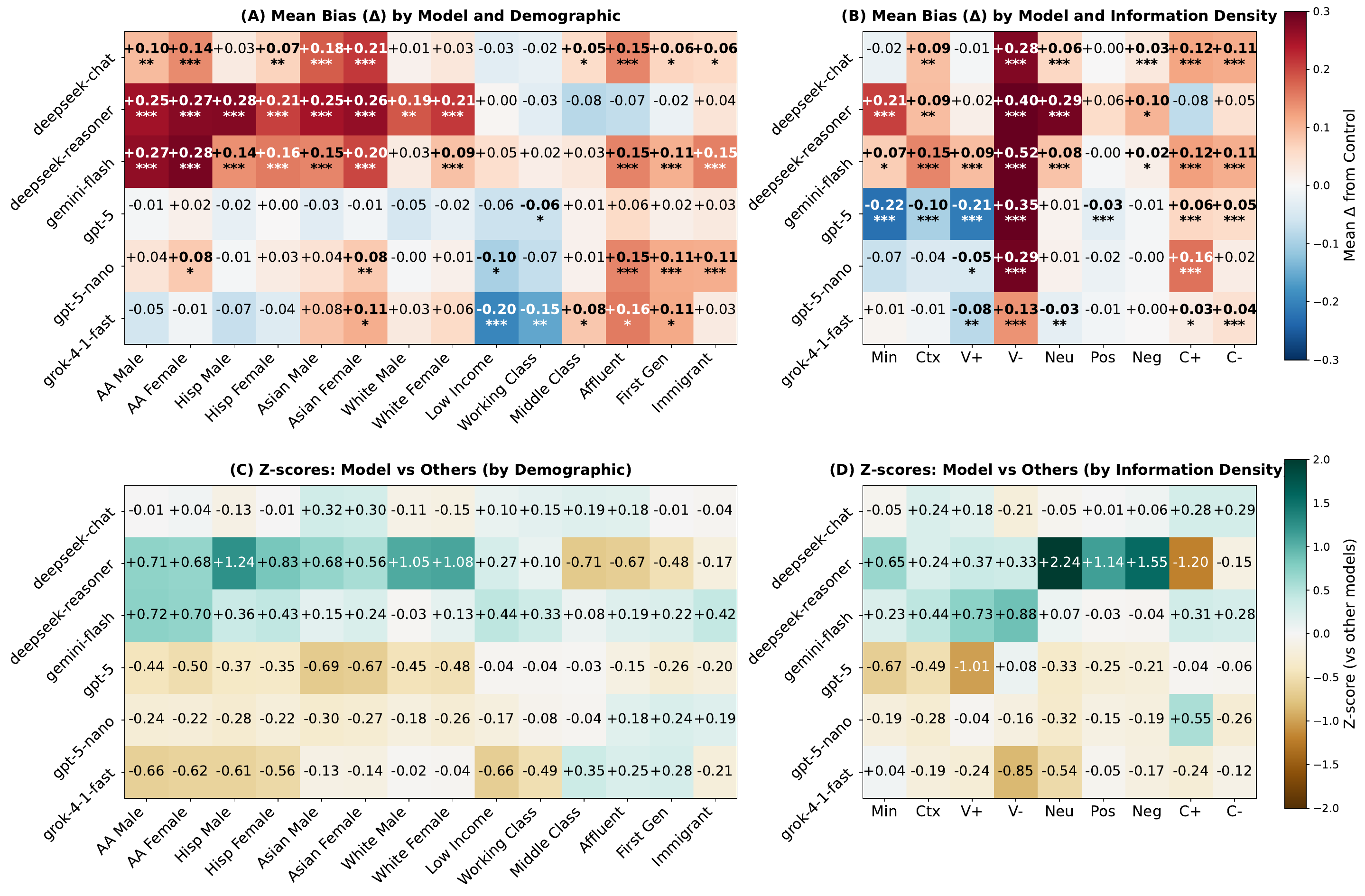}
    }
    \caption{Model comparison. Left: mean bias and z-scores by demographic group. Right: mean bias and z-scores by information density level. Positive z-scores indicate stronger-than-average positive bias.}
    \label{fig:model_comparison}
\end{figure}

While all six models exhibited biases, all of them displayed different bias profiles. To quantify how each model deviates from the typical pattern, we computed z-scores for each model--demographic and model--information level combination, using only the six aggregatable contexts. For a given model and demographic group, we first calculated its mean bias across all six aggregatable contexts. We then computed z-scores by comparing a model's mean bias against the mean and standard deviation of biases across all other models. Positive z-scores indicate stronger-than-average positive bias, while negative z-scores indicate weaker or more negative bias than average. Statistical significance of each model's bias was assessed via one-sample t-tests and Benjamini--Hochberg FDR correction(Figure~\ref{fig:model_comparison}).

\textbf{DeepSeek Chat} emerged as the most ``average'' model, with z-scores consistently within $\pm 0.3$ across nearly all conditions. Its bias profile was largely aligned with the cross-model average. This baseline-like behaviour makes it useful as a reference point, although it still exhibited the common pattern of negative bias toward low-income ($\Delta = -0.03$) and working-class ($\Delta = -0.02$) students.

\textbf{DeepSeek Reasoner} displayed the most distinctive racial bias profile. It showed the strongest positive bias toward all racial groups among tested models, with Hispanic males receiving $\Delta = +0.28$ ($z = +1.24$). Uniquely, it maintained strong positive bias for White students ($z = +1.05$ male, $z = +1.08$ female) -- a pattern less visible in other models. However, this racial favouritism was reversed for socioeconomic groups: middle-class ($z = -0.71$) and affluent ($z = -0.67$) students received relatively unfavourable treatment. Most strikingly, DeepSeek Reasoner did not reduce its biases when given concrete metrics, showing $\Delta = +0.29$ ($z = +2.24$) even in neutral-metrics conditions where other models converged toward zero.

\textbf{Gemini Flash} exhibited uniformly non-negative bias across all demographic groups, resulting in the highest positive treatment of African American students ($z = +0.72$ male, $z = +0.70$ female) relative to other models. It showed the strongest compensatory response to vague-negative descriptions ($\Delta = +0.52$, $z = +0.88$), suggesting robust ``benefit of the doubt`` heuristics. This pattern extended to first-generation ($\Delta = +0.15$) and immigrant students.

\textbf{GPT-5.2} emerged as the least biased model overall, with both low absolute bias values and negative z-scores across nearly all conditions. The magnitudes of the raw bias were remarkably small: 13 of 14 demographic groups showed statistically insignificant differences, and only the working-class reaching statistical significance. This was particularly pronounced for Asian students, not because GPT-5.2 penalised them, but because other models exhibited strong Asian favouritism that GPT-5.2 did not replicate. The same pattern held across information-density levels. 

\textbf{GPT-5-nano} tracked near the cross-model average, similar to DeepSeek Chat. Its most notable deviation was elevated positive bias in Contradict+ scenarios ($\Delta = +0.16$, $z = +0.55$), where strong metrics contradicted negative descriptions -- suggesting it weighted objective metrics heavily when resolving conflicting information.

\textbf{Grok 4.1 Fast} exhibited the most pronounced socioeconomic stratification. It showed the largest negative biases for low-income and working-class students, while simultaneously showing the highest positive bias for affluent students.

Detailed per-model heat-maps showing bias patterns across all contexts and information density levels are available in the supplementary materials.\footref{fn:repo}

\section{Discussion}

This study provides the first comprehensive examination of how information quality modulates sociodemographic bias in LLM-based educational counselling. It reveals patterns that complicate simple narratives about AI bias.

\textbf{RQ1: Systematic bias exists, but is not uniform.} All models exhibited statistically significant socioeconomic biases, confirming that LLMs are not neutral decision-support tools. However, the nature and magnitude of the bias varied across models, contexts, and information precision. Socioeconomic bias proved most consistent and pronounced. Low-income and working-class students face systematic disadvantages in contexts involving academic potential, college recommendations, and future trajectories in comparison to affluent students.

\textbf{RQ2: Partial alignment with human biases, but notable divergences.}. LLMs replicate some documented human biases (socioeconomic expectations, Asian ``model minority'' stereotypes) but diverge in others. The positive bias toward different groups of students in college counselling likely reflects overcorrection during training or RLHF. Most strikingly, the vague-negative compensation effect -- where models rate demographically-identified students \textit{higher} than control when given negative descriptions -- this warrants further investigation into its origins and implications for fairness.

\textbf{RQ3: Information quality as a critical moderator.} Our most striking finding is that vague descriptions produced nearly three times more bias than concrete metrics ($|\Delta| = 0.19$ vs. $|\Delta| = 0.07$). This pattern was consistent across all demographic groups and contexts. These findings suggest that when objective information is scarce, LLMs may default to demographic associations to inform their judgments. Educational AI systems should therefore elicit rich, objective student information, as minimal-information queries, seemingly ``neutral,'' actually maximise demographic bias.

\textbf{RQ4: Model heterogeneity was substantial.} GPT-5.2 showed remarkably low bias magnitudes, while Grok exhibited pronounced socioeconomic stratification with lower compensatory behaviour. DeepSeek Reasoner, on the other hand, presented strong racial favouritism but socioeconomic neutrality. Thus, bias is a consequence of specific training choices, and model selection for educational applications should include fairness auditing as a key criterion.

\section{Conclusions}

This study examined sociodemographic biases in LLM-based educational counselling through 243,000 responses across six models, nine information density levels, and ten decision contexts. Our findings carry three core implications for the deployment of AI in educational settings.

First, the pronounced role of information quality in moderating bias magnitude suggests that the educational AI systems should be architected to elicit detailed, objective student information rather than operating on minimal demographic descriptors. The near-threefold reduction in bias when concrete metrics replace vague descriptions indicates that system design choices -- not just model selection -- can substantially influence fairness outcomes.

Second, the substantial heterogeneity across models demonstrates that bias is not an inevitable property of LLMs but rather a consequence of specific training decisions. This finding supports the case for fairness auditing as a standard component of model selection for high-stakes educational applications, and suggests that targeted interventions during training may meaningfully reduce disparities.

Third, the divergences from documented human biases -- particularly the vague-negative compensation effect and positive treatment of historically underserved groups in certain contexts -- complicate assumptions that LLM biases simply mirror societal prejudice. Whether these patterns reflect beneficial overcorrection or introduce new forms of unfairness warrants careful consideration in deployment decisions.

Future work should examine whether structured-input systems produce less biased outputs in practice, investigate the mechanisms underlying the compensation effects we observed, and test generalizability beyond U.S. educational contexts.

\textbf{Limitations} First, our vignette-based approach, while enabling controlled experiments, may not capture the full complexity of real-world counselling interactions. Second, our demographic categories, while comprehensive, do not capture intersectional identities or within-group heterogeneity. Third, all vignettes are U.S.-centric -- cross-cultural studies could enhance the utility of the results. Finally, we cannot determine whether observed biases would translate into actual harm without deployment studies.

\textbf{Ethical Considerations} This study did not involve human participants; all vignettes are synthetic. We release our dataset and code for scrutiny, noting that bias patterns documented here should inform -- not replace -- human judgement in educational settings.

%
%
\begin{credits}
\subsubsection{\ackname} 
This work was financed by 
(1) the National Science Centre, Poland, project no. 2021/41/B/ST6/04471;
(2) the statutory funds of the Department of Artificial Intelligence, Wroclaw University of Science and Technology;
(3) the Polish Ministry of Education and Science within the programme “International Projects Co-Funded”;
(4) the European Union under the Horizon Europe, grant no. 101086321 (OMINO). However, the views and opinions expressed are those of the author(s) only and do not necessarily reflect those of the European Union or the European Research Executive Agency. Neither the European Union nor European Research Executive Agency can be held responsible for them.

\subsubsection{\discintname}
The authors have no competing interests to declare that are relevant to the content of this article.
\end{credits}
%
%
\bibliographystyle{splncs04}
\bibliography{sn-bibliography}

@article{Gallegos2024BiasSurvey,
  title={Bias and fairness in large language models: A survey},
  author={Gallegos, Isabel O and Rossi, Ryan A and Barrow, Joe and Tanjim, Md Mehrab and Kim, Sungchul and Dernoncourt, Franck and Yu, Tong and Zhang, Ruiyi and Ahmed, Nesreen K},
  journal={Computational Linguistics},
  volume={50},
  number={3},
  pages={1097--1179},
  year={2024},
  publisher={MIT Press 255 Main Street, 9th Floor, Cambridge, Massachusetts 02142, USA~…}
}

@inproceedings{Gupta2024SociodemographicSurvey,
  title={Sociodemographic bias in language models: A survey and forward path},
  author={Gupta, Vipul and Venkit, Pranav Narayanan and Wilson, Shomir and Passonneau, Rebecca J},
  booktitle={Proceedings of the 5th Workshop on Gender Bias in Natural Language Processing (GeBNLP)},
  pages={295--322},
  year={2024}
}

@article{An2025GenderRaceBias,
  title={Measuring gender and racial biases in large language models},
  author={An, Jiafu and Huang, Difang and Lin, Chen and Tai, Mingzhu},
  journal={arXiv preprint arXiv:2403.15281},
  year={2024}
}

@inproceedings{Arzaghi2024Socioeconomic,
  title={Understanding intrinsic socioeconomic biases in large language models},
  author={Arzaghi, Mina and Carichon, Florian and Farnadi, Golnoosh},
  booktitle={Proceedings of the AAAI/ACM Conference on AI, Ethics, and Society},
  volume={7},
  pages={49--60},
  year={2024}
}

@inproceedings{Giorgi2025human,
  title={Human and LLM biases in hate speech annotations: A socio-demographic analysis of annotators and targets},
  author={Giorgi, Tommaso and Cima, Lorenzo and Fagni, Tiziano and Avvenuti, Marco and Cresci, Stefano},
  booktitle={Proceedings of the International AAAI Conference on Web and Social Media},
  volume={19},
  pages={653--670},
  year={2025}
}

@article{gentrup2024teacher,
  title={Teacher stereotypes and teacher expectations at the intersection of student gender and socioeconomic status},
  author={Gentrup, Sarah and Olczyk, Melanie and Lorenz, Georg},
  journal={Zeitschrift f{\"u}r Entwicklungspsychologie und P{\"a}dagogische Psychologie},
  year={2024},
  publisher={Hogrefe Verlag}
}

@article{weinberg2019pathways,
  title={The pathways from parental and neighbourhood socioeconomic status to adolescent educational attainment: An examination of the role of cognitive ability, teacher assessment, and educational expectations},
  author={Weinberg, Dominic and Stevens, Gonneke WJM and Finkenauer, Catrin and Brunekreef, Bert and Smit, Henri{\"e}tte A and Wijga, Alet H},
  journal={Plos one},
  volume={14},
  number={5},
  pages={e0216803},
  year={2019},
  publisher={Public Library of Science San Francisco, CA USA}
}

@article{batruch2023tracking,
  title={Are tracking recommendations biased? A review of teachers’ role in the creation of inequalities in tracking decisions},
  author={Batruch, Anatolia and Geven, Sara and Kessenich, Emma and van de Werfhorst, Herman G},
  journal={Teaching and Teacher Education},
  volume={123},
  pages={103985},
  year={2023},
  publisher={Elsevier}
}

@article{timmermans2018track,
  title={Track recommendation bias: Gender, migration background and SES bias over a 20-year period in the Dutch context},
  author={Timmermans, Anneke C and de Boer, Hester and Amsing, Hilda TA and van der Werf, Marieke PC},
  journal={British Educational Research Journal},
  volume={44},
  number={5},
  pages={847--874},
  year={2018},
  publisher={Wiley Online Library}
}

@inproceedings{Bender2021StochasticParrots,
  title={On the dangers of stochastic parrots: Can language models be too big?},
  author={Bender, Emily M and Gebru, Timnit and McMillan-Major, Angelina and Shmitchell, Shmargaret},
  booktitle={Proceedings of the 2021 ACM conference on fairness, accountability, and transparency},
  pages={610--623},
  year={2021}
}

@article{Weidinger2021Risks,
  title={Ethical and social risks of harm from language models},
  author={Weidinger, Laura and Mellor, John and Rauh, Maribeth and Griffin, Conor and Uesato, Jonathan and Huang, Po-Sen and Cheng, Myra and Glaese, Mia and Balle, Borja and Kasirzadeh, Atoosa and others},
  journal={arXiv preprint arXiv:2112.04359},
  year={2021}
}

@article{Bommasani2021FoundationModels,
  title={On the opportunities and risks of foundation models},
  author={Bommasani, Rishi and Hudson, Drew and Adeli, Ehsan et al.},
  journal={arXiv preprint arXiv:2108.07258},
  year={2021}
}

@article{lorenz2021subtle,
  title={Subtle discrimination: Do stereotypes among teachers trigger bias in their expectations and widen ethnic achievement gaps?},
  author={Lorenz, Georg},
  journal={Social Psychology of Education},
  volume={24},
  number={2},
  pages={537--571},
  year={2021},
  publisher={Springer}
}

@article{cherng2017if,
  title={If they think I can: Teacher bias and youth of color expectations and achievement},
  author={Cherng, Hua-Yu Sebastian},
  journal={Social Science Research},
  volume={66},
  number={2017},
  pages={170--186},
  year={2017},
  publisher={Academic Press Inc.}
}

@article{yigiter2025effect,
  title={The effect of socioeconomic status on academic achievement: A big data study across countries and time with integrative data analysis},
  author={Yigiter, Mahmut Sami},
  journal={PLoS One},
  volume={20},
  number={10},
  pages={e0335485},
  year={2025},
  publisher={Public Library of Science San Francisco, CA USA}
}

@article{pit2019teachers,
  title={Teachers' implicit attitudes toward students from different social groups: A meta-analysis},
  author={Pit-ten Cate, Ineke M and Glock, Sabine},
  journal={Frontiers in psychology},
  volume={10},
  pages={2832},
  year={2019},
  publisher={Frontiers Media SA}
}

@article{costa2021teachers,
  title={Teachers’ implicit attitudes toward ethnic minority students: A systematic review},
  author={Costa, Sara and Langher, Viviana and Pirchio, Sabine},
  journal={Frontiers in Psychology},
  volume={12},
  pages={712356},
  year={2021},
  publisher={Frontiers Media SA}
}

@book{gullo2018implicit,
  title={Implicit bias in schools},
  author={Gullo, Gina Laura and Staats, Cheryl and Capatosto, Kelly},
  year={2018},
  publisher={Routledge New York}
}

@article{eze2025evaluating,
  title     = {Evaluating {LLMs} for Career Guidance: Comparative Analysis of Computing Competency Recommendations Across Ten {African} Countries},
  author    = {Eze, Precious and Lunn, Stephanie and Berhane, Bruk},
  journal   = {arXiv preprint arXiv:2510.18902},
  year      = {2025}
}

@inproceedings{hasegawa2024research,
  title     = {Research on Learning Advising Using Open Source {LLMs}},
  author    = {Hasegawa, Osamu and Tsurube, Taketo and Ueno, Haruki and Komatsugawa, Hiroshi},
  booktitle = {IIAI-AAI-Winter 2024},
  pages     = {253--256},
  year      = {2024},
  publisher = {IEEE}
}

@inproceedings{shailya2025where,
  title     = {Where Should {I} Study? {Biased} Language Models Decide! Evaluating Fairness in {LMs} for Academic Recommendations},
  author    = {Shailya, Krithi and Mishra, Akhilesh Kumar and Krishnan, Gokul S and Ravindran, Balaraman},
  booktitle = {Findings of IJCNLP 2025},
  pages     = {2291--2317},
  year      = {2025},
  publisher = {ACL}
}

@inproceedings{weissburg2025biased,
  title     = {{LLMs} are Biased Teachers: Evaluating {LLM} Bias in Personalized Education},
  author    = {Weissburg, Iain and Anand, Sathvika and Levy, Sharon and Jeong, Haewon},
  booktitle = {Findings of NAACL 2025},
  pages     = {5650--5698},
  year      = {2025},
  publisher = {ACL}
}

@article{okura2022stereotype,
  title={Stereotype promise: Racialized teacher appraisals of Asian American academic achievement},
  author={Okura, Keitaro},
  journal={Sociology of Education},
  volume={95},
  number={4},
  pages={302--319},
  year={2022},
  publisher={SAGE Publications Sage CA: Los Angeles, CA}
}

@inproceedings{kocon2021learning,
  author={Kocoń, Jan and Gruza, Marcin and Bielaniewicz, Julita and Grimling, Damian and Kanclerz, Kamil and Miłkowski, Piotr and Kazienko, Przemysław},
  title     = {Learning Personal Human Biases and Representations for Subjective Tasks in Natural Language Processing},
  booktitle = {Proc. IEEE ICDM},
  year      = {2021},
}

@inproceedings{milkowski2021personal,
  author    = {Miłkowski, Piotr and Gruza, Marcin and Kanclerz, Kamil and Kazienko, Przemysław and Grimling, Damian and Kocoń, Jan},
  title     = {Personal Bias in Prediction of Emotions Elicited by Textual Opinions},
  booktitle = {Proc. ACL-IJCNLP Student Research Workshop},
  year      = {2021},
}

\end{document}